\documentclass[12pt,a4paper]{scrartcl}

\usepackage{url}
\usepackage[english,french]{babel}
\usepackage[T1]{fontenc}
\usepackage{amsfonts}
\usepackage{fancyhdr}
\usepackage{graphics}
\usepackage{graphicx}
\usepackage{multicol}
\usepackage{lettrine}
\usepackage{hyperref}
\usepackage{url}
\usepackage{fancyhdr}
\usepackage[numbers]{natbib}

\title{Une représentation en graphe\\ pour l'enseignement de XML}

\date{}

\author{
     Emmanuel Desmontils\\
     \small{LINA, 2 rue de la Houssini\`ere}\\
     \small{BP92208, 44322 Nantes Cedex 03}\\
     \small{emmanuel.desmontils@univ-nantes.fr}
}

\newcommand{\MotsCles}[1]{\par\noindent 
{\small{\em Mots-clés\/}: #1}}

\newcommand{\Keywords}[1]{\par\noindent 
{\small{\em Keywords\/}: #1}}

\newcommand{\DTD}[1]{{\usefont{T1}{pcr}{l}{n}\small #1}}

\begin{document}

\onecolumn

\selectlanguage{french}

\maketitle

\begin{abstract}
XML est un format actuellement très utilisé. Dans le cadre des formations en informatique, il est indispensable d'initier les étudiants à ce format et, surtout, à tout son éco-système. Nous avons donc mis au point un modèle permettant d'appuyer l'enseignement de XML. Ce modèle propose de représenter un schéma XML sous la forme d'un graphe mettant en valeur les caractéristiques structurelles des documents valides. Nous présentons dans ce rapport les différents éléments graphique du modèle et les améliorations qu'il apporte à la modélisation de données en XML.
\newline
\MotsCles{XML, Représentation graphique, Schéma, DTD, XSD, Relax NG. }
\end{abstract}

\newpage

\tableofcontents                  

\newpage

\listoffigures  

\listoftables   


\newpage

\onecolumn


\section{Motivation et objectifs}

De nos jours, XML~\cite{world1998extensible,bray1997extensible} prend une place importante dans les systèmes informatiques. Ce format se retrouve aussi bien pour l'échange de données entre Web services, que pour le paramétrage d'applications ou pour mémoriser de façon pérenne des informations (à travers des bases de données XML~\cite{bourret1999xml} par exemple).

La complexité des documents est extrêmement variable. Même si beaucoup de structures XML (appelés schémas) sont simples, il est important de bien maîtriser la modélisation de tels documents. Il est souvent utile de s'appuyer sur des méthodologies de conceptions connues comme UML ou Merise~\cite{routledge2002uml,carlson2001modeling,lonjon2006modelisation,edModelisationXML}. Cependant, ces méthodes ne sont pas vraiment satisfaisantes pour le cas des données hiérarchiques.

De plus, pour exploiter ou produire des documents XML, un développeur doit être capable de bien appréhender les schémas. Cela lui permet de tirer profit au mieux de la structure hiérarchique à travers les API dédiées ou les langages adaptés. Nous avons constaté par ailleurs que les utilisateurs de XML n'exploitent pas toujours bien la structure hiérarchique de ce format. Il est donc important d'introduire, dans la formation des développeurs, un enseignement sur XML et son éco-système, comme en particulier :
\begin{itemize}
\item les différents langages de schéma (DTD, XSD~\cite{part20010}, Relax NG~\cite{clark2001relax}),
\item les API de programmation (SAX~\cite{megginson2001sax}, DOM~\cite{wood1998document}, etc.), 
\item les langages de requête et les bases de données XML (XPath~\cite{world1999xml,clark1999xml,fernandez2002xquery}, XQuery~\cite{world1999xml,fernandez2002xquery}, eXist~\cite{meier2003exist}, etc.), 
\item les langages de transformation (XSLT~\cite{transformationversion,clark1999xsl}).
\end{itemize}

Durant les nombreuses années de formation à XML, nous avons constaté que, pour tous ces outils, le choix d'une représentation graphique permet de mieux appréhender la structure du document à exploiter ou à produire.  Nous avons donc recherché une représentation sous forme de graphe pour mettre en évidence les caractéristiques structurelles du document à valider
. Nous nous sommes inspirés des outils graphiques utilisés pour représenter les modèles relationnels.

Dans le suite, nous allons présenter des exemples utilisant uniquement le langage de schéma originel pour XML : DTD. Cependant, l'utilisation de XSD ou de Relax NG serait identique. En effet, nous attachons le plus d'importance à l'aspect structurel des schémas plutôt qu'au typage des informations.
 
Après avoir présenté notre exemple fil rouge (section~\ref{section-exemple}) et les modèles existants (section~\ref{section-modeles}), nous présenterons notre modélisation graphique pour les éléments (section~\ref{section-element}), les attributs (section~\ref{section-attribut}) et la structuration (section~\ref{section-liens}). Après une discussion sur ce modèle (section~\ref{section-discussion}), nous conclurons. 

\section{Notre exemple}\label{section-exemple}

Afin d'illustrer notre modélisation, nous avons repris le sujet d'un exercice donné aux étudiants du Master MIAGE de Nantes en octobre 2013. Il concerne la modélisation d'un service (simplifié) de films à la demande auquel est adossé un réseau social spécialisé. Le texte décrivant le contexte, ainsi que la DTD associée sont donnés en annexe~\ref{annexe-sujet} et \ref{annexe-dtd}. Le modèle complet qui peut être produit à partir de ce schéma est donné en annexe~\ref{annexe-graphe}. 

\section{Modèles existants}\label{section-modeles}

\begin{figure*}[htb!]
\begin{center}
a) \includegraphics[scale=.33]{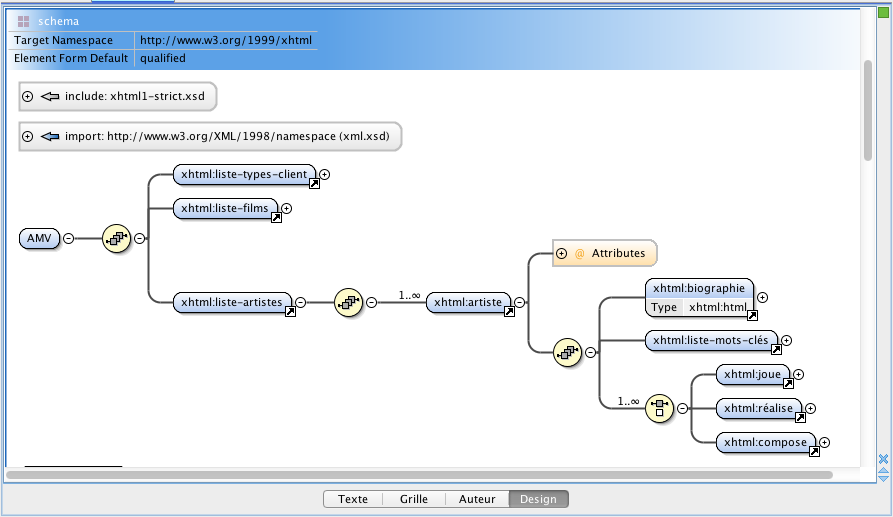} \\
b) \includegraphics[scale=.5]{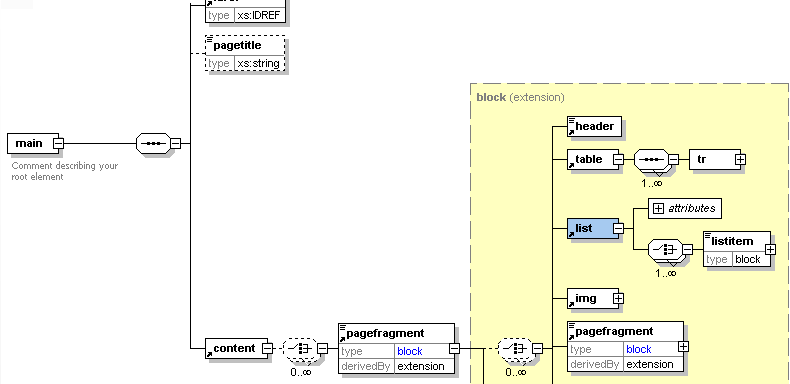}
\caption{Visualisation XSD avec (a) oXygen et (b) XMLSpy}
\label{fig-ex-xsd}
\end{center}
\end{figure*}

Les outils de manipulation dédiés à XML proposent couramment une représentation graphique pour les schémas XSD ou Relax NG. Les figures~\ref{fig-ex-xsd}a et \ref{fig-ex-xsd}b illustrent les représentations graphiques des schémas XSD\footnote{D'autres exemples de représentations graphiques de XSD peuvent être trouvées sur \url{http://en.wikipedia.org/wiki/XML_Schema_Editor}.} avec respectivement oXygen\footnote{\url{http://www.oxygenxml.com/xml_editor/xml_schema_editor.html} } et XMLSpy\footnote{\url{http://www.altova.com/xmlspy/xml-schema-editor.html}}. La représentation graphique de Relax NG (quand elle existe) est quasiment identique. Ces représentations ont en commun une représentation sous forme d'arbre (ou de forêt d'arbres) avec la possibilité de déployer ou non certaines branches. La structure en graphe n'est donc pas clairement visible et les liens de composition des éléments ne sont pas très lisibles.

Pour les DTD, il n'existe pas de représentation graphique dédiée pour ce type de schéma. 

Du point de vue général, il existe de nombreux outils graphiques de représentation des modèles conceptuels de données (diagrammes de classes pour UML~\cite{booch1998unified}, Entité-Association-Propriété pour la méthode Merise~\cite{tardieu1983methode,quang1991merise}, etc.). Ces modèles permettent de modéliser n'importe quelle structure de données, mais, de ce fait, ne sont pas nécessairement adaptés pour une compréhension des modèles hiérarchiques. 

\begin{figure*}[htb!]
\begin{center}
\includegraphics[scale=.75]{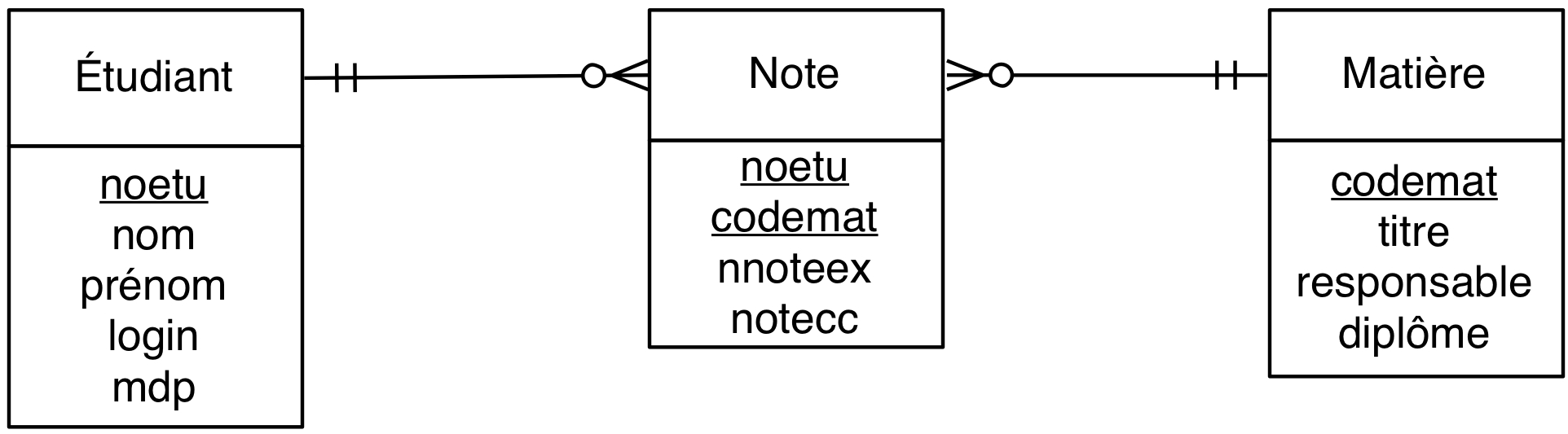} 
\caption{Exemple de "Crow's Foot Diagram" pour le modèle relationnel}
\label{fig-ex-crowsfoot}
\end{center}
\end{figure*}

Cependant, parmi ces modèles standard, nous nous sommes intéressés aux "Crow's Foot diagrams" de G. C. Everest~\cite{everest1976basic}. Ils sont utilisés par beaucoup de logiciels pour la représentation des tables et de leurs liens dans le modèle Entité-Relation\footnote{\url{http://en.wikipedia.org/wiki/Entity\%E2\%80\%93relationship_model}}. 
La figure~\ref{fig-ex-crowsfoot} présente un exemple d'utilisation de ce type de schéma pour la représentation d'un modèle relationnel. La forme graphique des cardinalités multiples (1..n ou 0..n) donne son nom à ce type de diagramme.

Notre objectif est donc de proposer une représentation graphique quelque soit le schéma XML (DTD, XSD, Relax NG, etc.) utilisant ce type de notation et permettant d'avoir une représentation visuelle donnant une intuition de la structure hiérarchique des documents XML valides.

\section{Modélisation des éléments}\label{section-element}

La notion d'élément XML est centrale, car elle amène le vocabulaire du dialecte et la grammaire. Dans cette première section de description du modèle, nous nous intéresserons uniquement à des éléments simples. Les éléments forment les sommets du graphe. Les liens entre les éléments correspondent aux contenus des éléments décrits par le schéma. Ces liens seront décrits en section~\ref{section-liens}.

\subsection{Élément vide}

\begin{figure}[htb!]
\begin{center}
\includegraphics[scale=1]{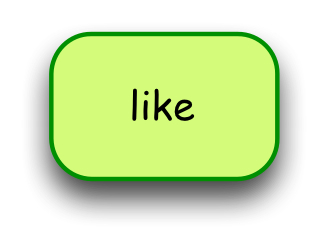} 
\caption{Élément vide}
\label{fig-ele-vide}
\end{center}
\end{figure}

Un élément est définit en DTD par \DTD{<!ELEMENT nom-élément contenu>}\footnote{Ici, comme dans la suite de ce document, les extraits de schéma en DTD seront présentés avec la police \DTD{Courrier}.} où "contenu" est une expression s'apparentant aux expressions régulières et qui décrit la structure du contenu de l'élément. Dans notre modèle, un élément est représenté par un rectangle vert contenant le nom de l'élément. La figure~\ref{fig-ele-vide} représente un élément vide décrit par \DTD{<!ELEMENT like EMPTY>}. 

\subsection{Élément à contenu textuel}

\begin{figure}[htb!]
\begin{center}
\includegraphics[scale=1]{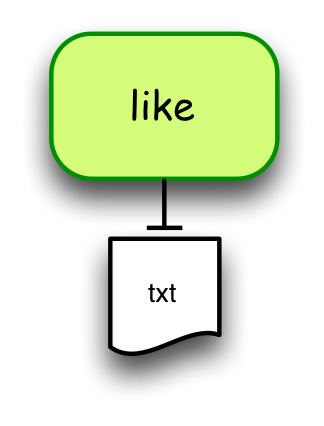} 
\caption{Élément à contenu textuel}
\label{fig-ele-txt}
\end{center}
\end{figure}

Certains éléments n'ont pas de contenu structuré : leur contenu est le plus souvent textuel. Dans notre modèle, les textes seront représentés par un rectangle blanc. Éventuellement, ce rectangle contiendra un terme décrivant le type de texte contenu. La figure~\ref{fig-ele-txt} représente un élément "like" avec un contenu textuel décrit en DTD par \DTD{<!ELEMENT like (\#PCDATA)>}.

Les éléments ne sont pas les seules sources d'information en XML : il y a aussi les attributs. Nous allons maintenant nous y intéresser.

\section{Modélisation des attributs}\label{section-attribut}

\subsection{Représentation d'un attribut}

XML autorise le positionnement d'attributs associés aux éléments. La liste des attributs est représentée par un rectangle jaune. Les attributs ont un type (CDATA, ID, etc.) et un comportement (\#REQUIRED, \#IMPLIED, etc.). Le tableau~\ref{tab-att} présente les cas d'attributs les plus fréquemment rencontrés et leur forme dans notre modèle.

\begin{table}[htd!]
\caption{Formes d'attributs}
\begin{center}
\begin{tabular}{|l|l|l|}
\hline
&{\bf Forme} &	{\bf Signification en DTD}\\
\hline
1&nom-cl &	\DTD{nom CDATA \#REQUIRED} \\ \hline
2&\%date-modif &	\DTD{date-modif CDATA \#IMPLIED} \\ \hline
3&pseudo &	\DTD{pseudo ID \#REQUIRED} \\ \hline
4&\#client &	\DTD{client IDREF \#REQUIRED} \\ \hline
5&\#(clients) &	\DTD{clients IDREFS \#REQUIRED} \\ \hline
6&{stars} & \DTD{stars (0|1|2|3|4) \#REQUIRED} \\ \hline
7&stars/'0' &	\DTD{stars CDATA '0'} \\
\hline
\end{tabular}
\end{center}
\label{tab-att}
\end{table}%

\begin{figure}[htb!]
\begin{center}
\includegraphics[scale=1]{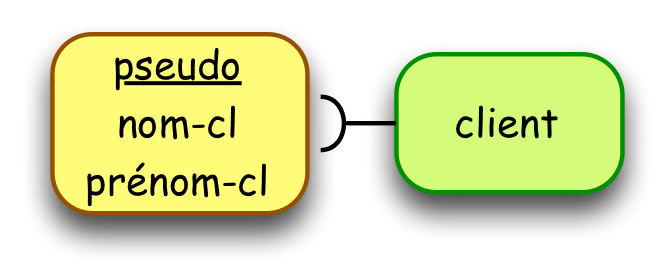} 
\caption{Attributs avec un identifiant}
\label{fig-ele-att1}
\end{center}
\end{figure}

Notons que, dans le cas d'une liste de valeurs possibles (ligne 6), notre modèle est incomplet. Ce n'est pas très grave au regard des objectifs de notre modélisation. Cependant, pour être plus complet, il est possible d'écrire "stars$\in$\{0,1,2,3,4,5\}". La figure~\ref{fig-ele-att1} représente par exemple un élément vide décrit en DTD par \DTD{<!ELEMENT client EMPTY>} avec trois attributs : 
\begin{itemize}
\item \DTD{<ATTLIST client pseudo ID \#REQUIRED>},  
\item \DTD{<ATTLIST client nom-cl CDATA \#REQUIRED>}, 
\item \DTD{<ATTLIST client prénom-cl CDATA \#REQUIRED>}.
\end{itemize}

\begin{figure}[htb!]
\begin{center}
\includegraphics[scale=1]{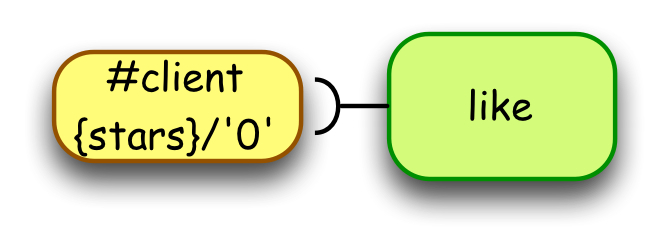} 
\caption{Attributs avec une référence}
\label{fig-ele-att2}
\end{center}
\end{figure}

Les cas du tableau~\ref{tab-att} peuvent être combinés. Par exemple, la figure~\ref{fig-ele-att2} propose un attribut "\{stars\}/'0'" qui est un attribut pris dans une liste de valeurs (par exemple "0|1|2|3|4|5") et avec comme valeur par défaut "0".

\subsection{Liens entre attributs}

\begin{figure}[htb!]
\begin{center}
\includegraphics[scale=1]{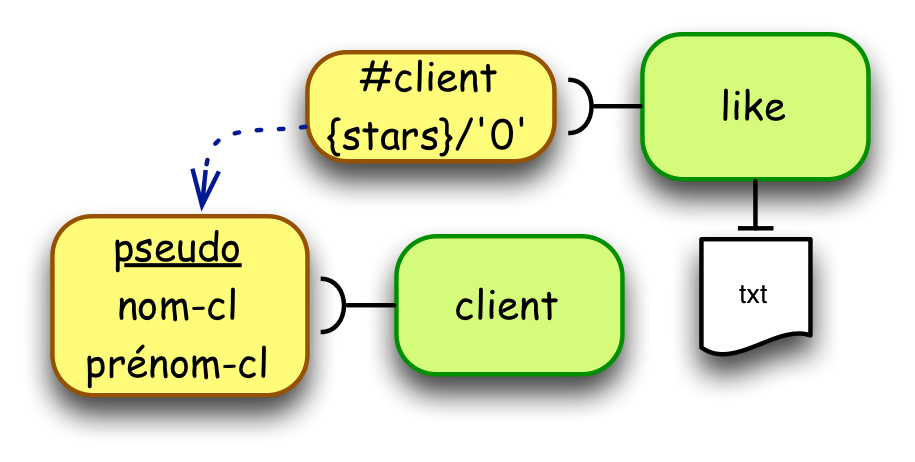} 
\caption{Lien entre ID et IDREF}
\label{fig-ele-liens-idref}
\end{center}
\end{figure}

Afin de préciser le rôle des attributs de type IDREF(S), il est possible d'ajouter une flèche en pointillé allant de l'attribut vers l'identifiant qu'il est supposé référencer. Les schémas XML ne prévoient pas de préciser ce lien, mais il facilite l'exploitation des documents DTD ou XSD (mise au point des API, utilisation des langages de recherche d'informations, etc.). La figure~\ref{fig-ele-liens-idref} montre le lien entre un attribut IDREF et l'attribut ID correspondant. Ici, l'attribut "client" de l'élément "like" doit contenir le "pseudo" d'un client.

\begin{figure*}[htb!]
\begin{center}
\includegraphics[scale=1]{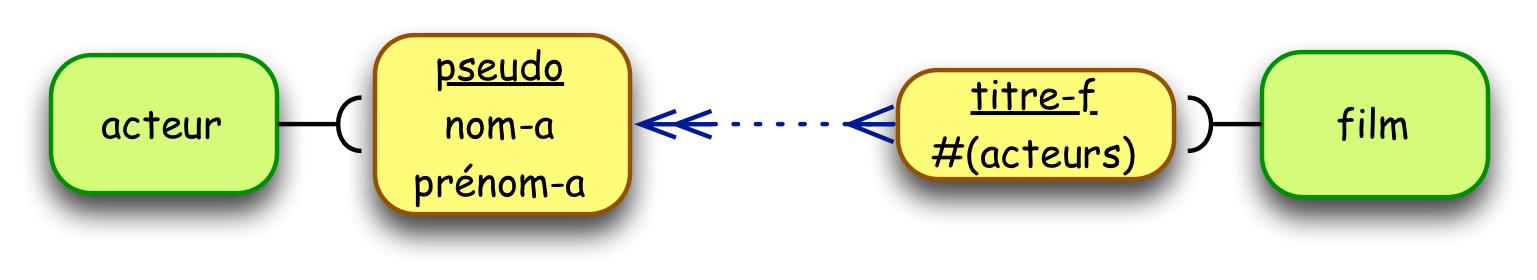} 
\caption{Lien entre ID et IDREFS}
\label{fig-ele-liens-idrefs}
\end{center}
\end{figure*}

La figure~\ref{fig-ele-liens-idrefs} représente le lien entre un attribut IDREFS (ici "acteurs") et l'attribut ID qui correspond ("pseudo").
 
Ces liens ne font pas à proprement parler partie du graphe. Ils sont plus des commentaires, des appuis, pour les  utilisateurs du graphe.

Maintenant que les noeuds du graphe ont été modélisés, nous allons nous intéresser à la modélisation des arcs.

\section{Modélisation des contenus complexes}\label{section-liens}

La structure de graphe n'apparaît que quand un élément en contient d'autres. Nous allons présenter les différents cas standard et élémentaires que nous pouvons rencontrer. 

\subsection{Itérations}\label{section-iteration}

\begin{figure*}[htb!]
\begin{center}
\includegraphics[scale=1]{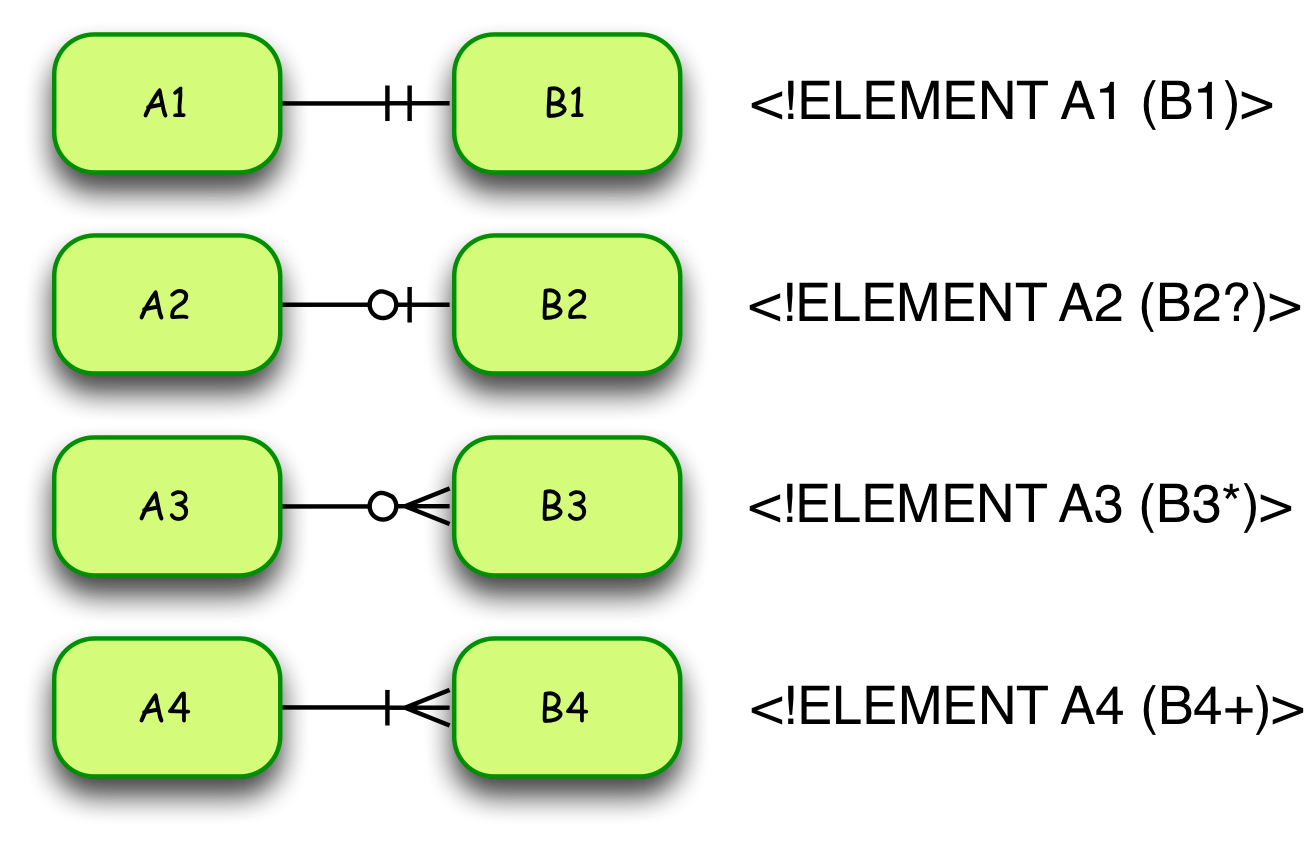}
\caption{Itération d'éléments}
\label{fig-ele-iteration}
\end{center}
\end{figure*}

Tout d'abord, nous allons représenter les opérateurs d'itération "*", "+" et "?". Pour cela, nous nous sommes inspirés des "Crow's foot diagrams" de G. C. Everest~\cite{everest1976basic} utilisés principalement pour représenter graphiquement les tables du modèle relationnel. La figure~\ref{fig-ele-iteration} présente la symbolique utilisée pour chacun des opérateurs. Ces opérateurs permettent de mettre en place les arcs élémentaires entre les différents sommets du graphe. 

\subsection{Séquences et alternatives}\label{section-seq-alt}

Il reste deux opérateurs à introduire : la composition d'éléments en séquence et l'alternative. Chacun des éléments qui composent la séquence ou l'alternative font l'objet d'un des opérateurs d'itération présentés dans la section~\ref{section-iteration}, d'une autre séquence, d'une autre alternative ou d'un sous-groupe présenté dans la section~\ref{section-ss-grp}.

\begin{figure}[htb!]
\begin{center}
\includegraphics[scale=1]{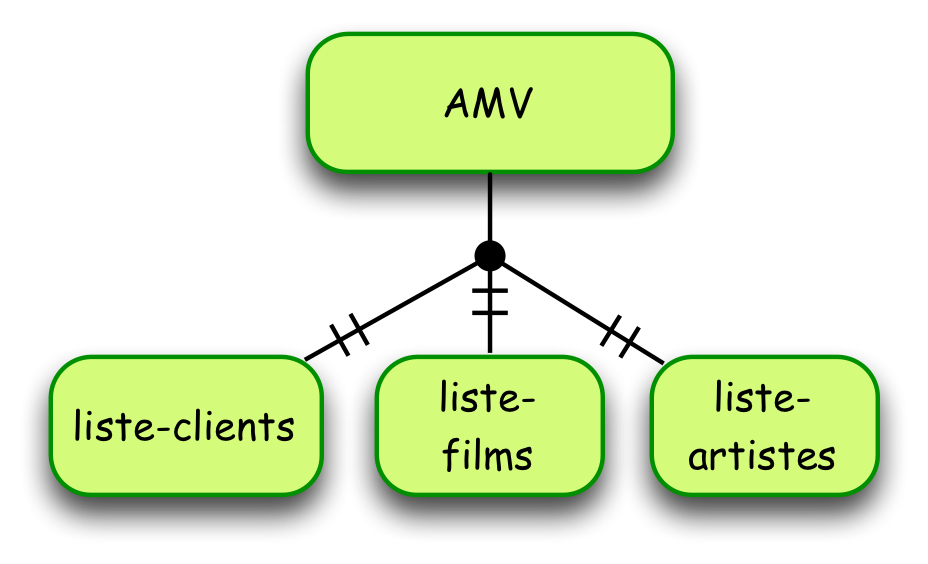} 
\caption{Composition d'éléments en séquence}
\label{fig-ele-seq}
\end{center}
\end{figure}

La séquence permet de décrire un contenu comme une succession d'éléments. Elle est représentée dans notre modèle par un point noir. L'ordre des n\oe uds est l'ordre dans le parcours trigonométrique autour de ce point en partant de la gauche du modèle.  La figure~\ref{fig-ele-seq} illustre une séquence et se traduit par \DTD{<!ELEMENT AMV (liste-clients, liste-films, liste-artistes)>}.

\begin{figure}[htb!]
\begin{center}
\includegraphics[scale=1]{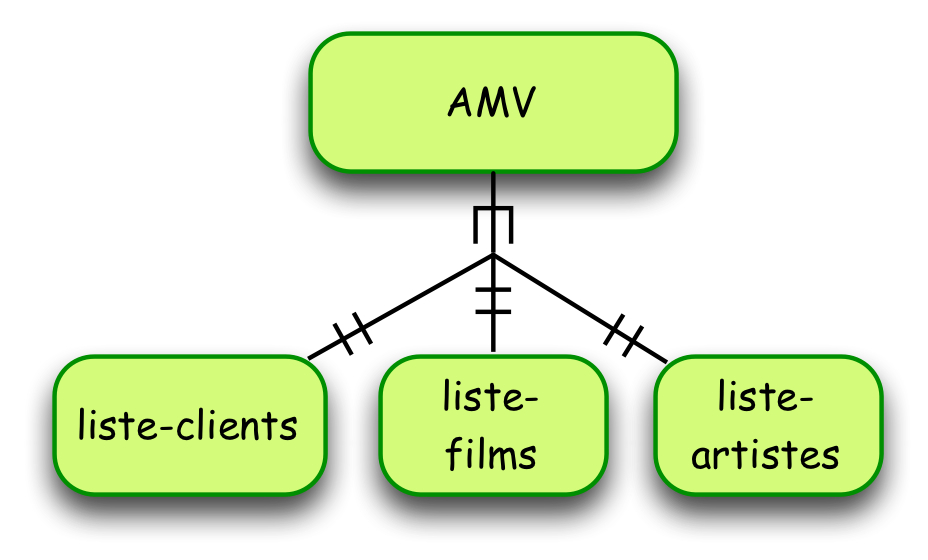}
\caption{Alternative d'éléments}
\label{fig-ele-alt}
\end{center}
\end{figure}

L'alternative permet de donner le choix entre plusieurs éléments. Elle est représentée dans notre modèle par une fourche. La figure~\ref{fig-ele-alt} illustre une alternative et se traduit par \DTD{<!ELEMENT AMV (liste-client | liste-films | liste-artistes)>}.

\subsection{Sous-groupes}\label{section-ss-grp}

Certaines descriptions de contenu sont construites en utilisant des sous-groupes. Par exemple, supposons qu'un artiste puisse avoir été acteur dans certains films, metteur en scène dans d'autres, voir compositeur de bande originale. Alors, nous pourrions proposer la définition suivante : \DTD{<!ELEMENT artiste (joue | réalise | compose)+>}. 

La partie \DTD{(joue | réalise | compose)} dans l'exemple ci-dessus est un sous-groupe sur lequel est appliqué l'opérateur d'itération "+" (vu dans la section~\ref{section-iteration}). Les trois éléments sont alors répétés dans un ordre quelconque (alternative vus dans la section~\ref{section-seq-alt}). 

\begin{figure}[htb!]
\begin{center}
\includegraphics[scale=1]{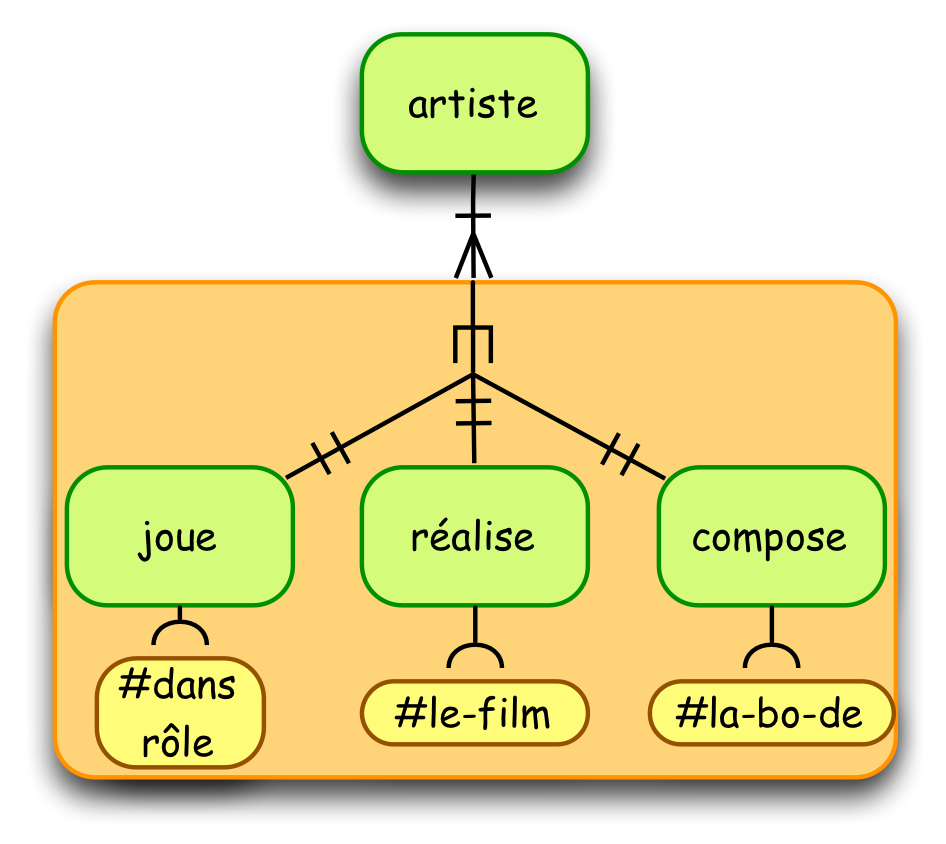}
\caption{Sous-groupe d'éléments}
\label{fig-ss-grp}
\end{center}
\end{figure}

Dans notre modèle, un sous-groupe est mis en évidence par une zone orangée. La figure~\ref{fig-ss-grp} illustre l'exemple ci-dessus. Cette notion de sous-groupe est "récursive" : un sous-groupe peut lui-même contenir des sous-groupes. Dans notre modèle, il y aura alors une imbrications de structures. 

\subsection{Schéma importé}

Pour terminer, il est parfois utile d'importer le schéma pré-existant d'un dialecte spécifique ou d'un langage standard. Dans ce cas, il n'est pas toujours utile de le détailler, car secondaire ou, au contraire, parfaitement maîtrisé. Dans ce contexte, il n'est pas nécessaire d'expliciter son graphe. Aussi, nous avons introduit dans notre modèle un symbole représentant un sous-graphe qui n'est pas détaillé. Pour cela, nous utilisons le symbole du nuage. 

\begin{figure}[htb!]
\begin{center}
\includegraphics[scale=1]{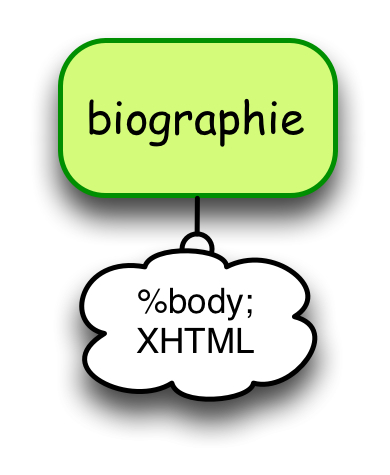}
\caption{Contenu d'un élément identique à celui de <body> en XHTML}
\label{fig-include}
\end{center}
\end{figure}

La figure~\ref{fig-include} représente la biographie d'un artiste en XHTML\footnote{La DTD de XHTML est décrite à l'URL : \url{http://www.w3.org/TR/xhtml1/DTD/xhtml1-strict.dtd}.}. Ce langage, bien connu, n'a pas besoin d'être représenté pour être manipulé. Il suffit alors de donner à "biographie" le même contenu que la balise "<body>" en XHTML (contenu qui est décrit par l'entité paramètre "\%body;").

\begin{figure}[htb!]
\begin{center}
\includegraphics[scale=1]{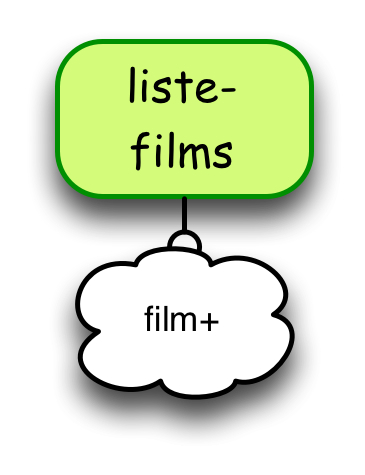} 
\caption{Contenu non-détaillé dans un élément}
\label{fig-hierarchique}
\end{center}
\end{figure}

Cette simplification peut aussi permettre de ne présenter qu'un graphe partiel. La partie mise en ellipse peut être considérée comme sous-entendue ou comme un "sous-graphe". Le graphe principal est alors considéré comme un graphe hiérarchique, la partie sous-entendue peut faire l'objet d'une graphe secondaire. La figure~\ref{fig-hierarchique} représente le contenu de l'élément "liste-films" comme une partie du graphe non développée.

\section{Discussion}\label{section-discussion}

\subsection{Les faiblesses du modèle}

Ce modèle a été mis au point à des fins pédagogiques. Il n'est donc pas adapté à des schémas très complexes que l'on peut trouver dans certains systèmes d'information. En effet, la topographie du graphe devient difficile à aborder sur une seule page A4 avec un grand nombre d'éléments ou lorsque le nombre des imbrications de groupes dépasse deux ou trois niveaux. Notre exemple complet en annexe~\ref{annexe-graphe} illustre bien cette idée. En effet, il ne comporte qu'une vingtaine d'éléments et semble, visuellement, déjà bien complexe. Par contre, il reste utilisable en conditions opérationnelles pour de petits schémas, souvent présents dans des applications Web à base de services.

Nous pouvons noter aussi que le graphe seul reste incomplet pour une vue exhaustive des éléments et des attributs. La description du contenu des attributs et des éléments reste très basique. Le modèle ne couvre même pas l'ensemble des types en DTD : les listes ne sont pas détaillées, les types NMTOKEN et NMTOKENS ne sont pas identifiés, tout comme le type NOTATION. Les entités générales ne sont pas non plus explicitées. Par contre, les entités paramètres le sont de manière indirecte (par leurs effets). Dans notre exemple en annexe~\ref{annexe-graphe}, le mécanisme d'inclusion de la DTD de XHTML s'effectue par ce principe. 

\subsection{Un modèle pédagogique}

Malgré tout, ces schémas restent un appui de taille pour aborder la construction de chemins de localisation en XPath, pour comprendre le mécanisme de parcours par l'API SAX, les optimisations de parcours avec DOM ou l'application des règles en XSLT. En XPath, par exemple, il est plus facile de construire le chemin pour atteindre une information en se référençant à notre modèle. En particulier, les liens entre les attributs de type IDREF ou IDREFS et leur correspondant de type ID permettent de mieux comprendre les fonctions XPath "id()" (en suivant la flèche) et "idref()" (en remontant le flèche). 

De plus, l'utilisation des symboles tirés des "Crow's foot diagrams" permet de donner une idée intuitive de la structure de l'arbre XML résultant. Pour renforcer l'intuition de l'arbre résultant, il est conseillé de travailler aussi sur la topographie du graphe. En effet, dans la plupart des cas, le graphe est déjà un arbre, voir un treillis (comme dans notre exemple en annexe~\ref{annexe-graphe}), et il est utile d'organiser sa structure comme tel. Ainsi, l'organisation globale et les "crow's foot" donnent une idée de ce que sera l'arbre XML résultant. 

\subsection{Un nouveau pas dans la modélisation XML}

\begin{figure*}[htb!]
\begin{center}
\includegraphics[scale=0.82]{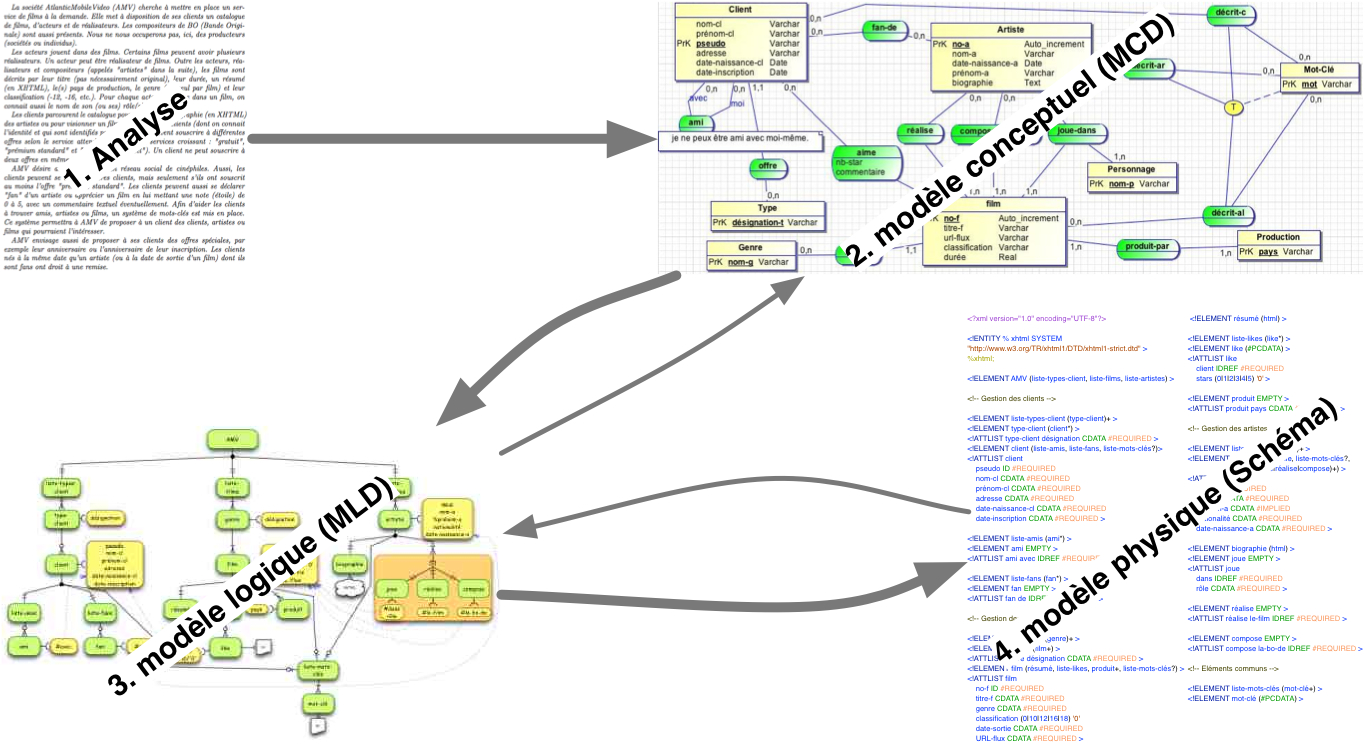}
\caption{Processus de modélisation avec notre modèle}
\label{fig-process}
\end{center}
\end{figure*}

Notre modélisation en graphe peut s'inscrire dans le processus de modélisation de données ou de connaissances sous forme hiérarchique, comme l'illustre la figure~\ref{fig-process}. 

En effet, un processus de modélisation standard commence par l'analyse du problème. Cette analyse amène la construction d'un modèle conceptuel des données, en Merise ou en UML par exemple, puis à un modèle physique, dans notre contexte en XML~\cite{routledge2002uml,carlson2001modeling,lonjon2006modelisation,edModelisationXML} (étapes 1, 2 et 4 dans la figure~\ref{fig-process}). Notre modèle peut s'insérer entre le modèle conceptuel et le schéma (étape 3 dans la figure~\ref{fig-process}). En effet, il peut être considéré comme le modèle logique des données, car il va mettre en valeur les propriétés du modèle physique XML (structure hiérarchique et contrôle syntaxique) et réaliser certains types de contraintes énoncées dans le modèle conceptuel. 

Ainsi, comme dans notre exemple en annexe~\ref{annexe-graphe}, il est possible de profiter de listes de valeurs courtes (les types d'abonnement par exemple) pour les transformer en contraintes structurelles. De la même manière, les contraintes du modèle conceptuel peuvent parfois être transformées en contraintes syntaxiques dans le modèle logique~\cite{edModelisationXML}. Dans notre exemple, la liste d'amis n'est possible que dans les abonnements "prémium".

Dans le cadre pédagogique uniquement, le processus de modélisation est alors pris à rebours en partant du schéma, objet de l'étude, pour remonter vers la modélisation en graphe pour mettre en évidence la structure.

\section{Conclusion}

Le modèle graphique de schémas XML que nous présentons ici est donc un modèle orienté vers l'aspect structurel du document XML et moins vers l'aspect type de données. C'est un modèle intéressant pour l'enseignement, surtout si le schéma n'est pas trop complexe. Il permet une bonne maîtrise du schéma, surtout pour la découverte de XPath, XSLT et des API de programmation (SAX et DOM). Il peut aussi être utilisé lors du processus de construction du schéma XML en jouant le rôle de modèle logique des données (entre le modèle conceptuel en Merise ou UML et le modèle physique qu'est XML).

Le modèle est opérationnel dans le cadre des enseignements de l'éco-système XML en Master à l'université de Nantes depuis quelques années. Malheureusement, actuellement, il n'existe pas encore d'outil logiciel permettant de gérer notre représentation, en particulier sa topographie, semi-automatiquement\footnote{L'ensemble des figures de ce rapport ont été conçues à l'aide du logiciel de dessin vectoriel OmniGraffle (\url{http://www.omnigroup.com/omnigraffle}).}.

\section{Remerciements}

Nous tenons à remercier les étudiants de la MIAGE de Nantes (en présentiel et à distance)\footnote{\url{http://miage.univ-nantes.fr/}} ainsi que les étudiants de Master CCI de Nantes\footnote{\url{http://www.dpt-info.univ-nantes.fr/44376489/0/fiche___pagelibre/}} d'avoir servi de cobaye pour ce travail pendant plusieurs années.

  \bibliographystyle{plainnat}
  \bibliography{bibliographie}

\appendix
\makeatletter
\def\@seccntformat#1{Annexe~\csname the#1\endcsname:\quad}
\makeatother


\section{Description de l'exemple}\label{annexe-sujet}

En octobre 2013, les étudiants de Master MIAGE 1ère année et de Master Informatique (spécialité ATAL) ont eu à modéliser en XML les données sur le projet suivant : 

\begin{quotation}\it
 La société AtlanticMobileVideo (AMV) cherche à mettre en place un service de films à la demande. Elle met à disposition de ses clients un catalogue de films, d'acteurs et de réalisateurs. Les compositeurs de BO (Bande Originale) sont aussi présents. Nous ne nous occuperons pas, ici, des producteurs (sociétés ou individus).
 
Les acteurs jouent dans des films. Certains films peuvent avoir plusieurs réalisateurs. Un acteur peut être réalisateur de films. Outre les acteurs, réalisateurs et compositeurs (appelés "artistes" dans la suite), les films sont décrits par leur titre (pas nécessairement original), leur durée, un résumé (en XHTML), le(s) pays de production, le genre (un seul par film) et leur classification (-12, -16, etc.). Pour chaque acteur qui joue dans un film, on connait aussi le nom de son (ou ses) rôle(s).

Les clients parcourent le catalogue pour consulter la biographie (en XHTML) des artistes  ou pour visionner un film particulier. Les clients (dont on connait l'identité et qui sont identifiés par un pseudo) peuvent souscrire à différentes offres selon le service attendu (par ordre de services croissant : "gratuit", "prémium standard" et "prémium universel"). Un client ne peut souscrire à deux offres en même temps.

AMV désire aussi constituer un réseau social de cinéphiles. Aussi, les clients peuvent :\begin{itemize}
\item se lier à d'autres clients s'ils ont souscrit au l'offre "prémium standard" ou "prémium universel" ; 
\item se déclarer "fan" d'un artiste ou apprécier un film en lui mettant une note (étoile) de 0 à 5, avec un commentaire textuel éventuellement, seulement s'ils ont souscrit l'offre "prémium universel".
\end{itemize}

Afin d'aider les clients à trouver amis, artistes ou films, un système de mots-clés est mis en place. Ce système permettra à AMV de proposer à un client des clients, artistes ou films qui pourraient l'intéresser.

AMV envisage aussi de proposer à ses clients des offres spéciales, par exemple leur anniversaire ou l'anniversaire de leur inscription. Les clients nés à la même date qu'un artiste (ou à la date de sortie d'un film) dont ils sont fans ont droit à une remise.
\end{quotation}

\section{Schéma DTD de l'exemple}\label{annexe-dtd}

On obtient alors le schéma suivant :
\begin{verbatim}
<?xml version="1.0" encoding="UTF-8"?>

<!ENTITY % xhtml 
                 SYSTEM 
                 "http://www.w3.org/TR/xhtml1/DTD/xhtml1-strict.dtd" >
%xhtml;

<!ELEMENT AMV (liste-types-client, liste-films, liste-artistes) >

<!-- Gestion des clients -->

<!ELEMENT liste-types-client (client)* >
<!ELEMENT client ((gratuit | prémium-standard | prémium-universel), 
                                liste-mots-clés?)>
<!ATTLIST client
    pseudo ID #REQUIRED
    nom-cl CDATA #REQUIRED
    prénom-cl CDATA #REQUIRED
    adresse CDATA #REQUIRED
    date-naissance-cl CDATA #REQUIRED
    date-inscription CDATA #REQUIRED >
    
<!ELEMENT gratuit EMPTY >
<!ELEMENT prémium-standard (liste-amis) >
<!ELEMENT prémium-universel (liste-amis, liste-fans) >

<!ELEMENT liste-amis (ami*) >
<!ELEMENT ami EMPTY >
<!ATTLIST ami avec IDREF #REQUIRED >

<!ELEMENT liste-fans (fan*) >
<!ELEMENT fan EMPTY >
<!ATTLIST fan de IDREF #REQUIRED >

<!-- Gestion des films -->

<!ELEMENT liste-films (genre)+ >
<!ELEMENT genre (film+) >
<!ATTLIST genre désignation CDATA #REQUIRED >
<!ELEMENT film (résumé, liste-likes, produit+, liste-mots-clés?) >
<!ATTLIST film
    no-f ID #REQUIRED
    titre-f CDATA #REQUIRED
    classification (0|10|12|16|18) '0'
    date-sortie CDATA #REQUIRED
    durée CDATA #REQUIRED
    URL-flux CDATA #REQUIRED >
    
<!ELEMENT résumé (%block;) > <!--  %block; contenu de <boby> -->

<!ELEMENT liste-likes (like*) >
<!ELEMENT like (#PCDATA) >
<!ATTLIST like
    client IDREF #REQUIRED
    date-like CDATA #REQUIRED
    stars (0|1|2|3|4|5) '0' >

<!ELEMENT produit EMPTY >
<!ATTLIST produit pays CDATA #REQUIRED >

<!-- Gestion des artistes -->

<!ELEMENT liste-artistes (artiste)+ >
<!ELEMENT artiste 
   (biographie, liste-mots-clés?, (joue|réalise|compose)+) >
<!ATTLIST artiste 
    no-a ID #REQUIRED
    nom-a CDATA #REQUIRED
    prénom-a CDATA #IMPLIED
    nationalité CDATA #REQUIRED
    date-naissance-a CDATA #REQUIRED >

<!ELEMENT biographie (%block;) >

<!ELEMENT joue EMPTY >    
<!ATTLIST joue
    dans IDREF #REQUIRED
    rôle CDATA #REQUIRED >
    
<!ELEMENT réalise EMPTY >
<!ATTLIST réalise le-film IDREF #REQUIRED >

<!ELEMENT compose EMPTY >
<!ATTLIST compose la-bo-de IDREF #REQUIRED >

<!-- Eléments communs -->

<!ELEMENT liste-mots-clés (mot-clé+) >
<!ELEMENT mot-clé (#PCDATA) >
\end{verbatim}

\onecolumn
\section{Exemple de graphe associé}\label{annexe-graphe}

\begin{figure*}[htb!]
\begin{center}
\includegraphics[angle=90,scale=0.59]{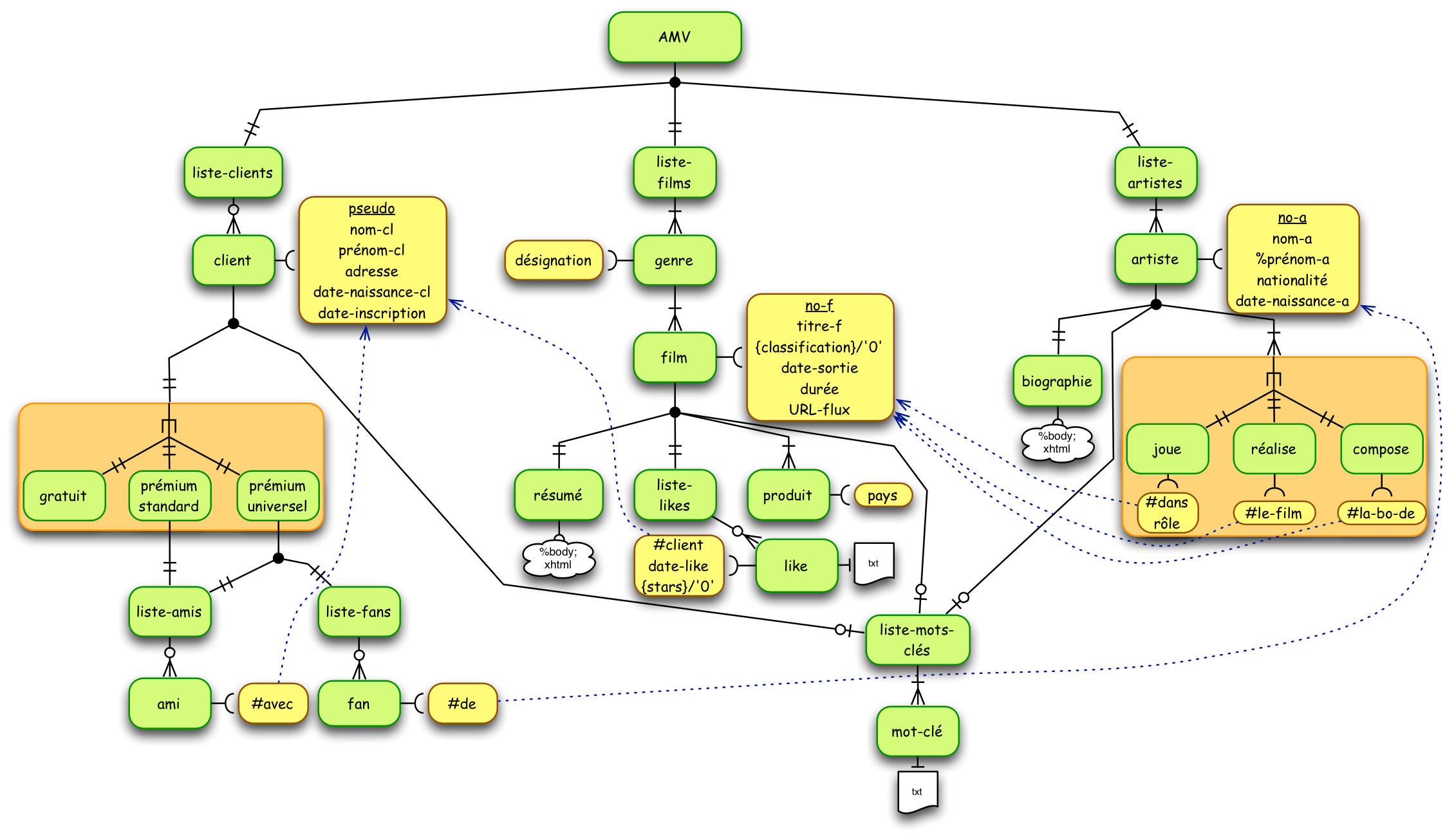}
\caption{AMV}
\label{fig-amv}
\end{center}
\end{figure*}

La figure~\ref{fig-amv} présente un exemple de graphe associé à notre exemple.

\newpage
\selectlanguage{english}

{\bf\large A Graph for the Learning of XML}

\begin{abstract} 
Currently, XML is a format widely used. In the context of computer science teaching, it is necessary to introduce students to this format and, especially, at its eco-system. We have developed a model to support the teaching of XML. We propose to represent an XML schema as a graph highlighting the structural characteristics of the valide documents. We present in this report different graphic elements of the model and the improvements it brings to data modeling in XML. 
\newline
\Keywords{XML, Schema, DTD, XSD, Relax NG, Graph.}
 \end{abstract}

\selectlanguage{french}  

\end{document}